\documentclass[twocolumn,prl,preprintnumbers,amsmath,amssymb,superscriptaddress]{revtex4}
\usepackage{graphicx}
\usepackage{dcolumn}
\usepackage{bm}
\usepackage{color} 

\usepackage{amsmath,amssymb}

\usepackage{fixmath}
\usepackage[normalem]{ulem}

\usepackage{comment}

\def\lesssim{\ \raise.3ex\hbox{$<$}\kern-0.8em\lower.7ex\hbox{$\sim$}\ }
\def\gesim{\ \raise.3ex\hbox{$>$}\kern-0.8em\lower.7ex\hbox{$\sim$}\ }

\newcommand \beq{\begin{eqnarray}}
\newcommand \eeq{\end{eqnarray}}

\usepackage[]{amsmath}
\usepackage{amssymb}
\usepackage{mathrsfs}
\usepackage[]{mathtools}
\usepackage[]{bm}
\usepackage[]{braket}
\usepackage{esint}
\usepackage{cases}

\usepackage{fontawesome}

\usepackage[]{graphicx}
\graphicspath{{fig/}}
\usepackage[caption=false]{subfig}

\usepackage[colorlinks=true]{hyperref}
\usepackage[]{comment}

\newcommand{\In}{\mathrm{in}}
\newcommand{\psil}{\psi_\mathrm{L}}
\newcommand{\psir}{\psi_\mathrm{R}}
\newcommand{\xv}{x_\mathrm{v}}

\begin{document}
\title{Multi-Particle Tunneling Transport at Strongly-Correlated Interfaces}

\author{Hiroyuki Tajima}
\affiliation{Department of Physics, School of Scinence, The University of Tokyo, Tokyo 113-0033, Japan}
\author{Daigo Oue}
\affiliation{The Blackett Laboratory, Department of Physics, Imperial Colledge London, Prince Consort Road, Kensington, London SW7 2AZ, United Kingdom}
\affiliation{%
Kavli Institute for Theoretical Sciences, University of Chinese Academy of Sciences, Beijing, 100190, China.
}%

\author{Mamoru Matsuo}
\affiliation{%
Kavli Institute for Theoretical Sciences, University of Chinese Academy of Sciences, Beijing, 100190, China.
}%

\affiliation{%
CAS Center for Excellence in Topological Quantum Computation, University of Chinese Academy of Sciences, Beijing 100190, China
}%
\affiliation{%
RIKEN Center for Emergent Matter Science (CEMS), Wako, Saitama 351-0198, Japan
}%
\affiliation{%
Advanced Science Research Center, Japan Atomic Energy Agency, Tokai, 319-1195, Japan
}%

\begin{abstract}
{We elucidate the multi-particle transport of pair- and spin-tunnelings in strongly correlated interfaces.}
Not only usual single-particle tunneling but also {interaction-induced} multi-particle tunneling processes naturally arise from a {conventional} microscopic model without any empirical parameters, 
through the overlap of the many-body wave functions around the {interface}.
We demonstrate how anomalous tunneling currents occur in a strongly interacting system due to the pair-tunneling process which we derived microscopically.
Our formulation is useful for junction systems in various disciplines, including atomtronics, spintronics, and nuclear reactions.
\end{abstract}
\maketitle
\noindent {\it Introduction.---}
{Transport phenomena are of tremendous interest as a probe of state of matter in modern physics. 
In the last century, various quantum many-body phenomena such as superconductivity~\cite{BCS} and Kondo effect~\cite{Kondo} were observed via transport measurements.
Recently, it is the most important problem to understand how transport phenomena reflect physical properties in strongly-correlated systems such as high-$T_{\rm c}$ superconductors~\cite{Pruschke} and dense quark matter~\cite{Alford}.}
\par
To understand an interplay between strong correlations and non-equilibrium properties,
it is crucial to develop a versatile framework for tunneling transport via an interface from microscopic arguments.
In particular,
a spin-exchange tunneling has been discussed widely in the context of spintronics~\cite{Konig2003,Adachi2011,Ohnuma2014,Ohnuma2017,Tatara2017,MM2018,Inoue2017,Kato2019,Silaev1,Silaev2,Ominato1,Ominato2,Kato2020,Yama2021,Yamamoto2021,Oue2021}, instead of a usual single-particle tunneling process.
Moreover, multi-particle tunnelings such as pair-transfer are known to play
a significant role in nuclear systems~\cite{Potelreview,Scamps,Magierski,Potel}.
However, the microscopic origin and mechanism of these nontrivial tunneling processes are not well understood despite their importance across research fields because of the complexities in the systems.
\par
{On the other hand, ultracold atomic gases provide favorable opportunities to study quantum many-body phenomena in a systematic way thanks to their cleanness, as called {\it atomtronics}~\cite{Amico}}
Various transport phenomena have been observed in
state-of-the-art experiments in such systems ~\cite{Stadler,Krinner2,Brown,Pace,Brantut,Husmann,Hausler,Valtolina2015,Kwon}.
One of the hottest recent topics in cold atoms is anomalous transport in strongly interacting Fermi gases where non-equilibrium many-body physics plays a crucial role~\cite{Chien,Krinner,Enss}.
\begin{figure}[t]
    \centering
    \includegraphics[width=8cm]{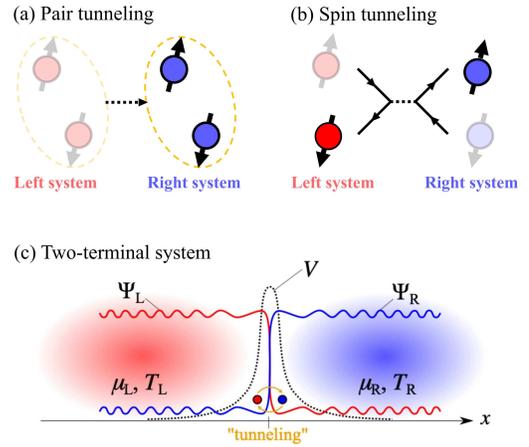}
    \caption{(a) pair and (b) spin tunnelings at the interface between strongly correlated systems.
    The panel (c) shows the two-terminal setup with a potential barrier $V$ (where we omit the spin-dependence for simplicity).
    We assume that local equilibrium is achieved in each terminal far from the potential barrier [i.e., $(\mu_{\rm L},T_{\rm L})\neq (\mu_{\rm R},T_{\rm R})$] and is described by asymptotic mode functions.
    Conversely, the many-body wave functions $\Psi_{\rm L,R}$ of the two reservoirs overlap around the potential barrier (e.g., the proximity effect), which induces tunneling processes based on the usual Hamiltonian with a two-body interaction shown in Eq.~(\ref{eq:Hr}).}
    \label{fig:1}
\end{figure}
\par
Considerable theoretical effort has been made to understand the anomalous tunneling transport observed in these experiments~\cite{Uchino2017,Liu2017,Yao2018,Han2019,Furutani2020,Sekino2020,Uchino20201,Uchino2020,Zaccanti,Piselli,Setiawan}. 
In particular, the pairing fluctuation effects 
have been discussed extensively for the Bardeen--Cooper--Schrieffer(BCS) to Bose--Einstein-condensation (BEC) crossover~\cite{Chen2005,Zwerger,Strinati,Ohashi2020}.
However, the detailed structure of interface
prevents the understanding of a microscopic tunneling process.
Accordingly, while a single-particle tunneling Hamiltonian with empirical parameters has been introduced in previous theoretical studies~\cite{Uchino2017,Liu2017,Yao2018,Han2019,Furutani2020,Sekino2020,Uchino20201,Uchino2020}, it is difficult to determine how many-body effects appear in the tunneling transport.
Furthermore, it is unclear how to realize the spin-exchange tunneling~\cite{Konig2003,Adachi2011,Ohnuma2014,Ohnuma2017,Tatara2017,MM2018,Inoue2017,Kato2019,Silaev1,Silaev2,Ominato1,Ominato2,Kato2020,Yama2021,Yamamoto2021,Oue2021} in atomtronic systems.

\par
In this study, we derive the tunneling Hamiltonian from a generic model.
Not only the usual one-body tunneling term but also the interaction-induced pair- and spin-exchange tunneling terms, which were overlooked in cold-atom communities, naturally arise from 
the overlap of wave functions (see Fig.~\ref{fig:1}).
{While the system
consisting of a potential barrier with an arbitrary form [e.g. a soft wall, as shown in Fig.~\ref{fig:1}(c)] and a two-body interaction
is relevant directly for cold atoms,
it is also useful to understand fundamental aspects of transport at interfaces between strongly-correlated systems.}
{Our formulation reveals a clear connection between cold atoms and other junction systems in the context of spintronics as well as nuclear reactions.}
For example, our derivation of the spin-tunneling process from the bulk Hamiltonian can be applied to the study of magnonic spin transport in normal-metal-ferromagnet junctions, where particles are immobile in each lattice site~\cite{MM2018}.
{Moreover, multi-neutron transfer in nuclei can be described by the pair-tunneling Hamiltonian derived in this study~\cite{Potelreview}.}
{Our formalism can easily be extended to the case with long-range interactions and three-body ones~\cite{SM},
which are of interdisciplinary interest in condensed matter~\cite{Hata} and nuclear systems~\cite{Hammer}.
}
\par
Using the momentum-space effective Hamiltonian combined with the Schwinger-Keldysh formalism~\cite{Rammer}, 
we demonstrate the conduction properties in strongly interacting spin-1/2 Fermi gases with a two-terminal setup and show that a non-trivial pair-tunneling current occurs even in the linear-response regime.
In particular, we demonstrate the temperature dependence of the ratio between the mass conductance and the Seebeck coefficient in the strong-coupling regime.
In the following, we take $\hbar=k_{\rm B}=1$.
\par

\noindent {\it Real-space Hamiltonian.---}
We start from non-relativistic spin-1/2 fermions (i.e., the pseudospin is given by $\sigma=\uparrow,\downarrow$) with the contact-type interaction described by~\cite{Zwerger}
\begin{align}
\label{eq:Hr}
    \hat{H}&=\int d^3\bm{r}\sum_{\sigma}\hat{\psi}_{\sigma}^\dag(\bm{r})\hat{h}_{\sigma}(\bm{r})\hat{\psi}_{\sigma}(\bm{r}) \cr
    &+g\int d^3\bm{r}\hat{\psi}_{\uparrow}^\dag(\bm{r})\hat{\psi}_{\downarrow}^\dag(\bm{r})\hat{\psi}_{\downarrow}(\bm{r})\hat{\psi}_{\uparrow}(\bm{r}),
\end{align}
where
\begin{align}
    \hat{h}_{\sigma}(\bm{r})=-\frac{\nabla^2}{2m_\sigma}+\hat{V}_\sigma(\bm{r})
\end{align}
is the one-body local Hamiltonian with the external potential barrier $\hat{V}_{\sigma}(\bm{r})$.
In Eq.~(\ref{eq:Hr}), $g$ is the two-body coupling constant.
One can rewrite $\hat{\psi}_{\sigma}(\bm{r})$ as
\begin{align}
\label{eq:psi_LR}
    \hat{\psi}_{\sigma}(\bm{r})=\hat{\psi}_{\sigma,{\rm L}}(\bm{r})+\hat{\psi}_{\sigma,{\rm R}}(\bm{r}).
\end{align}
Because we are interested in the low-energy tunneling current between two reservoirs ${\rm L}$ and ${\rm R}$ in thermal equilibrium, we assume that $\hat{\psi}_{\sigma,{\rm i}}(\bm{r})$ denotes the field operator for wave functions with biased probability distributions to the ${\rm i} \in \{{\rm L,R}\}$ reservoirs (noting that these wave functions have an overlap around the potential barrier).
{We specify $\hat{\psi}_{\rm L,R}$ when moving to the level (momentum) space,
based on one-particle scattering theory as described later}.
Using $\hat{\psi}_{\sigma,{\rm L,R}}(\bm{r})$, we rewrite $\hat{H}$ as
\begin{align}
    \hat{H}=\hat{H}_{\rm L}+\hat{H}_{\rm R}+\hat{H}_{\rm 1r}+\hat{H}_{\rm 1t}+\hat{H}_{\rm 2t}+\hat{H}_{\rm ind.},
    \label{eq:H}
\end{align}
where the reservoir Hamiltonian is
\begin{align}
\label{eq:H_LR}
    \hat{H}_{{\rm i} ={\rm L,R}}&=
    \int d^3\bm{r}\sum_{\sigma}\hat{\psi}_{\sigma,{\rm i}}^\dag(\bm{r})
    \left[-\frac{\nabla^2}{2m_\sigma}\right]
    \hat{\psi}_{\sigma,{\rm i}}(\bm{r})\cr
    &+g\int d^3\bm{r}\hat{\psi}_{\uparrow,{\rm i}}^\dag(\bm{r})\hat{\psi}_{\downarrow,{\rm i}}^\dag(\bm{r})\hat{\psi}_{\downarrow,{\rm i}}(\bm{r})\hat{\psi}_{\uparrow,{\rm i}}(\bm{r}),
\end{align}
the one-body reflection term is
\begin{align}
    \hat{H}_{\rm 1r}=\int d^3\bm{r}\sum_{\sigma,{\rm i}}\hat{\psi}_{\sigma,{\rm i}}^\dag(\bm{r})\hat{V}_\sigma(\bm{r})\hat{\psi}_{\sigma,{\rm i}}(\bm{r}),
    \label{eq:H_1r}
\end{align}
and the one-body tunneling term is
\begin{align}
\label{eq:H_1t}
    \hat{H}_{\rm 1t}=\int d^3\bm{r}\sum_{\sigma}\left[\hat{\psi}_{\sigma,{\rm L}}^\dag(\bm{r})\hat{\tau}_\sigma(\bm{r})\hat{\psi}_{\sigma,{\rm R}}(\bm{r})+{\rm h.c.}\right].
\end{align}
In Eq.~(\ref{eq:H_1t}), we introduced $\hat{\tau}_{\sigma}(\bm{r})=\hat{h}_{\sigma}(\bm{r})+g\sum_{{\rm i}}\hat{N}_{\bar{\sigma},{\rm i}}(\bm{r})$ 
with the density operator $\hat{N}_{\sigma,{\rm i}}(\bm{r})=\hat{\psi}_{\sigma,{\rm i}}^\dag(\bm{r})\hat{\psi}_{\sigma,{\rm i}}(\bm{r})$ ($\bar{\sigma}$ denotes the opposite spin of $\sigma$).
Note that Eq.~(\ref{eq:H_1t}) indicates the tunneling and reflection processes occur locally.
This fact reflects that the two wave functions in the ${\rm L}$ and ${\rm R}$ states are not completely separated but overlap as a result of the proximity effect.
Therefore, tunneling occurs via the overlap of the many-body wave functions $\Psi_{\rm L}$ and $\Psi_{\rm R}$ near the potential barrier in our model (see Fig.~\ref{fig:1}).

Similarly, we obtain the interaction-induced tunneling term
$\hat{H}_{\rm 2t}=\hat{H}_{\rm pair}+\hat{H}_{\rm spin}$ consisting of
the pair tunneling
\begin{align}
    \hat{H}_{\rm pair}=g\int d^3\bm{r}\left[\hat{P}_{\rm L}^\dag(\bm{r})\hat{P}_{\rm R}(\bm{r})+{\rm h.c.}\right]
\end{align}
with the pair creation operator $\hat{P}_{\rm i}^\dag(\bm{r})=\hat{\psi}_{\uparrow,{\rm i}}^\dag(\bm{r})\hat{\psi}_{\downarrow,{\rm i}}^\dag(\bm{r})$ and
the spin-exchange tunneling
\begin{align}
    \hat{H}_{\rm spin}&=g\int d^3\bm{r} \left[\hat{S}_{\rm L}^{+}(\bm{r})\hat{S}_{\rm R}^{-}(\bm{r})+\hat{S}_{\rm R}^{+}(\bm{r})\hat{S}_{\rm L}^{-}(\bm{r})\right]
\end{align}
with the spin ladder operators $\hat{S}_{\rm i}^{+}(\bm{r})=\hat{\psi}_{\uparrow,{\rm i}}^\dag(\bm{r})\hat{\psi}_{\downarrow,{\rm i}}(\bm{r})$ and $\hat{S}_{{\rm i}}^{-}(\bm{r})=\hat{\psi}_{\downarrow,{\rm i}}^\dag(\bm{r})\hat{\psi}_{\uparrow,{\rm i}}(\bm{r})$.
Note that a similar two-body tunneling has been discussed in the numerical simulation of a few-body system with a double-well trap potential~\cite{Erdmann}.
Also, the induced interface interaction $\hat{H}_{\rm ind.}$ between two systems reads
\begin{align}
    \hat{H}_{\rm ind.}=g\int d^3\bm{r} \sum_{\sigma}\hat{N}_{\sigma,{\rm L}}(\bm{r})\hat{N}_{\bar{\sigma},{\rm R}}(\bm{r}).
    \label{eq:H_ind}
\end{align}
We emphasize that the derivation starting from the model with the contact-type interaction does not involve any approximations.
\par
\noindent {\it Momentum-space effective Hamiltonian.---}
Because the potential barrier peaks around the center of the system ($x=0$) and vanishes at $x\rightarrow\pm\infty$,
there are free particles in the far region.
Here, we approximately evaluate Eqs.~(\ref{eq:H_LR})--(\ref{eq:H_ind}) by substituting the asymptotic form of the wave functions given by,
\begin{align}
\psi_{\sigma,\mathrm{L}}(\bm{r})
&= \sum_{\bm{k}} 
\widetilde{c}_{\bm{k},\sigma,{\rm L}} \times 
\begin{cases}
e^{i\bm{k}\cdot\bm{r}}+A_{\bm{k},\sigma} e^{-i\bm{k}\cdot\bm{r}}
& (x < 0),
\\
B_{\bm{k},\sigma} e^{i\bm{k}\cdot\bm{r}}  & (x > 0),
\end{cases}
\label{eq:L_asymptotic}
\\
\psi_{\sigma,\mathrm{R}}(\bm{r})
&= 
\sum_{\bm{k}}
\widetilde{c}_{\bm{k},\sigma,{\rm R}} \times 
\begin{cases}
B_{\bm{k},\sigma} e^{-i\bm{k}\cdot\bm{r}}
& (x < 0),
\\
e^{-i\bm{k}\cdot\bm{r}}+A_{\bm{k},\sigma} e^{i\bm{k}\cdot\bm{r}}
 & (x > 0),
\end{cases}
\label{eq:R_asymptotic}
\end{align}
where $\widetilde{c}_{\bm{k},\sigma,\mathrm{i}}$ is the amplitude of the asymptotic wave function~\cite{SM}.
Note that $A_{\bm{k},\sigma}$  and $B_{\bm{k},\sigma}$ are c-numbers corresponding to the reflection and transmission coefficients.
Assuming that the tunneling effect thorough the one-body potential is weak,
we substitute Eqs.~(\ref{eq:L_asymptotic}) and (\ref{eq:R_asymptotic}) into Eq.~(\ref{eq:H_LR}) to obtain the effective reservoir Hamiltonian:
\begin{align}
    H_{{\rm i}={\rm L,R}}&=\sum_{\bm{k},\sigma,{\rm i}}\varepsilon_{\bm{k},\sigma}c_{\bm{k},\sigma,{\rm i}}^\dag c_{\bm{k},\sigma,{\rm i}}\cr
    &+g\sum_{\bm{k},\bm{k}'\bm{q}}
    c_{\bm{k}+\bm{q},\uparrow,{\rm i}}^\dag
    c_{-\bm{k},\downarrow,{\rm i}}^\dag
    c_{-\bm{k}',\downarrow,{\rm i}}
    c_{\bm{k}'+\bm{q},\uparrow,{\rm i}},
\end{align}
where the amplitude $\widetilde{c}_{\bm{k},\sigma,{\rm i}}$ is replaced with a fermionic annihilation operator $c_{\bm{k},\sigma,{\rm i}}$ and the kinetic energy is defined as $\varepsilon_{\bm{p},\sigma}=p^2/(2m_\sigma)$.
Similarly,
substituting Eqs.~(\ref{eq:L_asymptotic}) and (\ref{eq:R_asymptotic}) into Eqs.~(\ref{eq:H_1r}) and (\ref{eq:H_1t}), we obtain the reflection and tunneling Hamiltonians in the momentum space:
\begin{align}
    H_{\rm 1r}&=\sum_{\bm{p},\bm{k},\sigma,{\rm i}}\mathcal{R}_{\bm{k},\bm{p},\sigma,i}c_{\bm{k},\sigma,{\rm i}}^\dag c_{\bm{p},\sigma,{\rm i}},
\\
    H_{\rm 1t}&=\sum_{\bm{p},\bm{k},\sigma}
    \mathcal{T}_{\bm{k},\bm{p},\sigma}
    \left[
    c_{\bm{k},\sigma,{\rm L}}^\dag c_{\bm{p},\sigma,{\rm R}}
    +
    c_{\bm{k},\sigma,{\rm R}}^\dag c_{\bm{p},\sigma,{\rm L}}
    \right],
\end{align}
where terms up to the first-order have been retained in the transmission coefficient.
The one-body tunneling amplitude is defined as
$\mathcal{T}_{\bm{k},\bm{p},\sigma}={Z_{\bm{k},\bm{p},\sigma,{\rm L,R}}}[\delta_{\bm{k},\bm{p}}\varepsilon_{\bm{p},\sigma}+V_{\sigma}(\bm{k}-\bm{p})+g\sum_{i}N_{\bm{k}-\bm{p},\bar{\sigma},{\rm i}}]$ and the reflection amplitude is defined as $\mathcal{R}_{\bm{k},\bm{p},\sigma,{\rm i}}={Z_{\bm{k},\bm{p},\sigma,{\rm i},{\rm i}}}V_{\sigma}(\bm{k}-\bm{p})$.
Note that we have defined the overlap integral between the two reservoirs
$Z_{\bm{k},\bm{p},\sigma,{\rm i},{\rm j}}=\int d\bm{r}f_{\bm{k},\sigma,{\rm i}}^*(\bm{r})f_{\bm{p},\sigma,{\rm j}}(\bm{r})$,
symbolically writing Eqs.~(\ref{eq:L_asymptotic}) and (\ref{eq:R_asymptotic}) as $\psi_{\sigma,{\rm i}}(\bm{r})=\sum_{\bm{k}}\tilde{c}_{\bm{k},\sigma,{\rm i}}f_{\bm{k},\sigma,{\rm i}}(\bm{r})$.
For the two-body term,
we use the relationship 
$\int d\bm{r}g\hat{\psi}_{Q_1}^\dag(\bm{r})\hat{\psi}_{Q_2}^\dag(\bm{r})\hat{\psi}_{Q_3}(\bm{r})\hat{\psi}_{Q_4}(\bm{r})
    =:\sum_{Q_{1\cdots4}}\tilde{g}_{Q_{1\cdots4}}c_{Q_1}^\dag
   c_{Q_2}^\dag c_{Q_3} c_{Q_4}$,
where $Q_\ell$ denotes the state label of each $\hat{\psi}$. 
At the leading order with respect to $A_{\bm{k},\sigma}$ and $B_{\bm{k},\sigma}$, one can obtain {the pair tunneling coupling
$\tilde{g}_{Q_{1\cdots4}}^{\rm (pair)}\simeq g\left(B_{\bm{k}_1,\sigma_1}^*B_{\bm{k}_2,\sigma_2}^*+B_{\bm{k}_3,\sigma_3}B_{\bm{k}_4,\sigma_4}\right)$
and the spin-tunneling one
$\tilde{g}_{Q_{1\cdots4}}^{\rm (spin)}\simeq g\left(B_{\bm{k}_1,\sigma_1}^*B_{\bm{k}_2,\sigma_2}+B_{\bm{k}_3,\sigma_3}^*B_{\bm{k}_4,\sigma_4}\right)$.}
Assuming the long-wavelength limit for the transmitted waves, namely, $B_{\bm{k},\sigma}\rightarrow B_{\bm{0},\sigma}$,
we obtain the interaction-induced tunneling terms as
\begin{align}
\label{eq:H_pair_FT}
    H_{\rm pair}&=
    \sum_{\bm{q}} {\mathcal{T}_{2,{\rm pair}}}
    \left[{P}_{\bm{q},{\rm L}}^\dag {P}_{-\bm{q},{\rm R}}+{\rm h.c.}\right],
    \\
\label{eq:H_spin_FT}
    H_{\rm spin}&=
    \sum_{\bm{q}} {\mathcal{T}_{2,{\rm spin}}}
    \left[{S}_{\bm{q},{\rm L}}^{+}{S}_{\bm{q},{\rm R}}^{-}+{S}_{\bm{q},{\rm R}}^{+}{S}_{\bm{q},{\rm L}}^{-}\right],
    \\
\label{eq:H_ind_FT}
    H_{\rm ind.}&=
    \sum_{\bm{q},\sigma}{\mathcal{T}_{2,{\rm ind.}}}
    {N}_{\bm{q},\sigma,{\rm L}}
    {N}_{\bm{q},\bar{\sigma},{\rm R}},
\end{align}
where we keep only the leading-order terms {and defined $\mathcal{T}_{2,{\rm pair}}=2g{\rm Re}[B_{\bm{0},\uparrow}B_{\bm{0},\downarrow}]$, 
$\mathcal{T}_{2,{\rm spin}}=2g{\rm Re}[B_{\bm{0},\uparrow}^*B_{\bm{0},\downarrow}]$,
and $\mathcal{T}_{2,{\rm ind.}}=g[B_{\bm{0},\uparrow}^*B_{\bm{0},\uparrow}+B_{\bm{0},\downarrow}^*B_{\bm{0},\downarrow}]$
}.
In Eqs.~(\ref{eq:H_pair_FT})--(\ref{eq:H_ind_FT}), ${X}_{\bm{q}}=\int d\bm{r}\hat{X}(\bm{r})e^{i\bm{q}\cdot\bm{r}}$ is the usual Fourier component of $\hat{X}(\bm{r})$.
We emphasize that our formulation covers the tunneling properties of not only one-body and pair transfers in the entire BCS--BEC crossover regime 
but also {the spin-tunneling process widely discussed in spintronics~\cite{Konig2003,Adachi2011,Ohnuma2014,Ohnuma2017,Tatara2017,MM2018,Inoue2017,Kato2019,Silaev1,Silaev2,Ominato1,Ominato2,Kato2020,Yama2021,Yamamoto2021,Oue2021}.
Note that $H_{\rm spin}$ is absent in the case with only interactions between identical fermions (e.g., $p$-wave Feshbach resonance)~\cite{SM}.
}

\noindent {\it Tunneling currents.---}
Let us derive the tunneling current formulas based on the momentum-space effective Hamiltonian.
The mass current operator is defined by
$I_M=i\left[N_{{\rm tot.},{\rm L}},H\right]$,
where $N_{{\rm tot.},{\rm L}}=\sum_{\sigma}N_{\bm{0},\sigma,{\rm L}}\equiv\sum_{\bm{k},\sigma}c_{\bm{k},\sigma,{\rm L}}^\dag c_{\bm{k},\sigma,{\rm L}}$.
Its explicit form is
\begin{align}
\label{eq:I}
    I_M&=i\sum_{\bm{p},\bm{k},\sigma}
    \mathcal{T}_{\bm{k},\bm{p},\sigma}\left[c_{\bm{k},\sigma,{\rm L}}^\dag c_{\bm{p},\sigma,{\rm R}}-c_{\bm{k},\sigma,{\rm R}}^\dag c_{\bm{p},\sigma,{\rm L}}\right]\cr
    &\quad+2i\sum_{\bm{q}}
    {\mathcal{T}_{2,{\rm pair}}}
    \left[P_{\bm{q},{\rm L}}^\dag P_{-\bm{q},{\rm R}}-P_{-\bm{q},{\rm R}}^\dag P_{\bm{q},{\rm L}}\right].
\end{align}
We are interested in the statistical average $\langle I_M \rangle={\rm Tr}\left[\rho I_M\right]$, 
where $\rho$ is the density matrix.
Using the Langreth rule with the truncation with respect to the tunnel couplings up to the second order~\cite{Langreth}, we obtain 
\begin{align}
\notag
&\langle I_{M}\rangle
=\langle I_M \rangle_{\rm 1t} 
    + \langle I_M \rangle_{\rm pair},
    \\ \notag
    &:=4\sum_{\bm{k},\bm{p},\sigma}\int \frac{d\omega}{2\pi}\mathcal{T}_{\bm{k},\bm{p},\sigma}^2{\rm Im}G_{\bm{k},\sigma,{\rm L},\omega}^{\rm ret.}{\rm Im}G_{\bm{p},\sigma,{\rm R},\omega}^{\rm ret.}
    \delta f_{\bm{k},\bm{p},\sigma,\omega}^{\rm neq.}
    \\ \label{eq:Ipair}
    &\hspace{.3em}
    +8\sum_{\bm{q}}\int \frac{d\omega}{2\pi}
    {\mathcal{T}_{2,{\rm pair}}^2}
    {\rm Im}\Gamma_{\bm{q},{\rm L},\omega}^{\rm ret.}{\rm Im}\Gamma_{\bm{q},{\rm R},\omega}^{\rm ret.}\delta b_{\bm{q},\omega}^{\rm neq.},
\end{align}
where $G_{\bm{k},\sigma,{\rm i},\omega}^{\rm ret.}$ and
$\Gamma_{\bm{q},{\rm i},\omega}^{\rm ret.}$
are the retarded components of the single-particle Green's function and pair susceptibility, respectively. 
The biases between the two reservoirs are incorporated into the non-equilibrium distribution differences $\delta f_{\bm{k},\bm{p},\sigma,\omega}^{\rm neq.}=f_{\bm{k},\sigma,{\rm L},\omega}-f_{\bm{p},\sigma,{\rm R},\omega}$
and $\delta b_{\bm{q},\omega}^{\rm neq.}=b_{\bm{q},{\rm L},\omega}-b_{\bm{q},{\rm R},\omega}$ with respect to the one- and two-particle states. 
Note that these distribution functions involving the local chemical potentials $\mu_{\rm i}$ and temperatures $T_{\rm i}$ can be obtained from the lesser propagators as
$f_{\bm{k},\sigma,{\rm i},\omega}=-G_{\bm{k},\sigma,{\rm i},\omega}^{<}/\left[2i{\rm Im}G_{\bm{k},\sigma,{\rm i},\omega}^{\rm ret.}\right]$
and
$b_{\bm{q},{\rm i},\omega}=\Gamma_{\bm{q},{\rm i},\omega}^{<}/\left[2i{\rm Im}\Gamma_{\bm{q},{\rm i},\omega}\right]$, respectively.
In the strong-attraction limit ($g<0$) of the normal phase ($a^{-1}\rightarrow +\infty$), one can assume an approximate form of the pair susceptibility as $\Gamma_{\bm{q},{\rm i},\omega}^{\rm ret.}\propto \left(\omega+i\delta-\frac{q^2}{4m}-\Sigma_{\bm{q},{\rm i},\omega}\right)^{-1}$~\cite{Pini2020}, which is nothing more than the retarded Green's function of a tightly bound molecule (where $\Sigma_{\bm{q},{\rm i},\omega}$ is the bosonic self-energy).
This fact combined with Eq.~(\ref{eq:Ipair}) indicates that the pair-tunneling current naturally appears in the strong-coupling regime within the linear response approach; meanwhile, such a pair transport proportional to $g^2$ does not appear in the non-interacting case as expected.
This is in sharp contrast with the previous studies, where non-linear tunneling currents~\cite{Uchino2017,Han2019,Furutani2020} and additional tunneling amplitudes with respect to closed-channel molecules~\cite{Liu2017} are considered to explain the anomalous transport induced by strong pairing fluctuations. 
\par
\begin{figure}
    \centering
    \includegraphics[width=7.5cm]{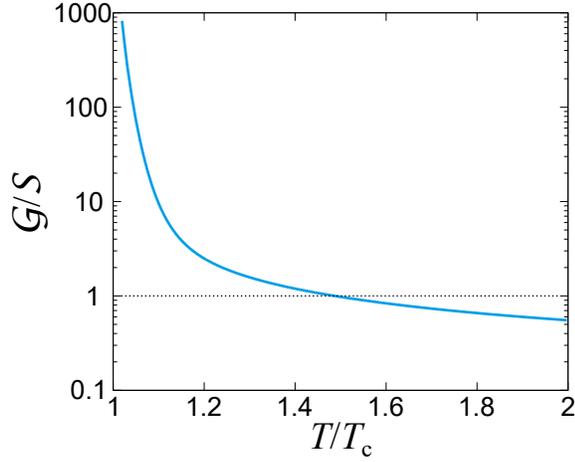}
    \caption{Ratio between the pair-tunneling-induced mass conductance $\mathcal{G}$ and Seebeck coefficient $\mathcal{S}$ as a function of temperature $T$ in the strong-coupling limit ($a^{-1}\rightarrow \infty$). $T_{\rm c}$ is the superfluid critical temperature.}
    \label{fig:2}
\end{figure}
To see the importance of $H_{\rm pair}$, in Fig.~\ref{fig:2}, we plot the ratio between the pair-tunneling mass conductance $\mathcal{G}=\lim_{\Delta\mu\rightarrow 0}\frac{\langle I_M\rangle_{\rm pair}}{\Delta \mu}$ and Seebeck coefficient
$\mathcal{S}=\lim_{\Delta T\rightarrow 0}\frac{\langle I_M\rangle_{\rm pair}}{\Delta T}$ in the strong-coupling limit ($a^{-1}\rightarrow \infty$) at low temperatures~\cite{SM,Watabe2021,Tajima2022},
where we defined $\Delta\mu=\mu_{\rm L}-\mu_{\rm R}$ and $\Delta T=T_{\rm L}-T_{\rm R}$; however, we eventually take $T_{\rm L}=T_{\rm R}\equiv T$ and $\mu_{\rm L}=\mu_{\rm R}\equiv \mu$, leading to $N_{\rm tot.{\rm L}}=N_{\rm tot.{\rm R}}\equiv N_{\rm tot.}$.
One can find a dramatic enhancement of $\mathcal{G}/\mathcal{S}$ near the molecular BEC temperature $T_{\rm c}\simeq 0.218T_{\rm F}$~\cite{Zwerger} where $T_{\rm F}=\frac{k_{\rm F}^2}{2m}$ is the Fermi energy of an ideal Fermi gas at zero temperature.
This result indicates that $\mathcal{G}$ is remarkably sensitive to changes in the molecular distribution $\delta b_{\bm{q},\omega}^{\rm neq.}$.
Indeed, near $T=T_{\rm c}$, the bosonic distribution function exhibits an infrared divergence as a result of the gapless excitation~\cite{Andersen}.
Because the shift of $\mu$ is directly relevant to the emergence of the gapless excitation, $\mathcal{G}$ tends to be larger than $\mathcal{S}$.
On the other hand, $\mathcal{S}$ becomes larger than $\mathcal{G}$ at high temperatures because the change of $T$ involves a large number of thermal excitations. 
Even though we consider the strong-coupling BEC limit,
the enhancement of pair-induced conduction can manifest
anomalous transport in a unitary Fermi gas~\cite{Husmann,Pace}.
A quantitative comparison with existing experiments
is left for future work.
Note that the single-particle tunneling current $\langle I_{M}\rangle_{\rm 1t}$ is strongly suppressed in the strong-coupling BEC regime at low temperatures, where most fermions in the system form tightly bound molecules (i.e., $\delta f_{\bm{k},\bm{p},\sigma,\omega}^{\rm neq.}\rightarrow 0$).
Because the system can be described by molecular bosons with one-boson tunneling, our approach below $T_{\rm c}$ would be consistent with the bosonic superfluid transport discussed in Ref.~\cite{Uchino2020}.
In such a case, the conduction of the condensed pair would also play an important role~\cite{Zaccanti,Piselli,Setiawan,Kanasz}.
At high temperature, $\langle I_{M}\rangle_{\rm 1t}$ becomes large as a result of the enhancement of $\mathcal{T}_{\bm{k},\bm{p},\sigma}$ associated with thermally excited fermions. 
\par

Moreover, 
$H_{\rm spin}$ induces the spin current operator $I_{S}=i\left[N_{\bm{0},\uparrow,{\rm L}}-N_{\bm{0},\downarrow,{\rm L}},H\right]$, such that
\begin{align}
\label{eq:Is}
I_S&=i\sum_{\bm{p},\bm{k},\sigma}\eta_{\sigma}
    \mathcal{T}_{\bm{k},\bm{p},\sigma}\left[c_{\bm{k},\sigma,{\rm L}}^\dag c_{\bm{p},\sigma,{\rm R}}-c_{\bm{k},\sigma,{\rm R}}^\dag c_{\bm{p},\sigma,{\rm L}}\right]\cr
    &\quad+2i\sum_{\bm{q}}
    {\mathcal{T}_{2,{\rm spin}}}
    \left[
S_{\bm{q},{\rm L}}^{+}S_{-\bm{q},{\rm R}}^{-}
-S_{\bm{q}{\rm L}}^{-}S_{-\bm{q},{\rm R}}^{+}
\right],
\end{align}
where we define $\eta_{\sigma}=\delta_{\sigma,\uparrow}-\delta_{\sigma,\downarrow}$.
In the presence of a spin-dependent bias,
we obtain the statistical average of the spin current,
\begin{align}
\notag
&\langle I_S\rangle
    =\langle I_S\rangle_{\rm 1t}+\langle I_S\rangle_{\rm spin},
    \\ \notag
    &:= 4\sum_{\bm{k},\bm{p},\sigma}\int \frac{d\omega}{2\pi}\mathcal{T}_{\bm{k},\bm{p},\sigma}^2{\rm Im}G_{\bm{k},\sigma,{\rm L},\omega}^{\rm ret.}{\rm Im}G_{\bm{p},\sigma,{\rm R},\omega}^{\rm ret.}
    \eta_{\sigma}\delta f_{\bm{k},\bm{p},\sigma,\omega}^{\rm neq.}
    \\
    &\hspace{.3em}
    +8\sum_{\bm{q}}
    \int\frac{d\omega}{2\pi}
    {\mathcal{T}_{2,{\rm spin}}^2}
    {\rm Im}\chi_{\bm{q},{\rm L},\omega}^{\rm ret.}
    {\rm Im}\chi_{\bm{q},{\rm R},\omega}^{\rm ret.}
    \delta \bar{b}_{\bm{q},\omega}^{\rm neq.},
\end{align}
where $\chi_{\bm{q},{\rm i},\omega}^{\rm ret.}$ is the retarded component of the dynamical spin susceptibility.
$\delta \bar{b}_{\bm{q},\omega}^{\rm neq.}=\bar{b}_{\bm{q},{\rm L},\omega}-\bar{b}_{\bm{q},{\rm R},\omega}$
is the non-equilibrium distribution difference of magnon-like excitations.
Using the lesser spin susceptibility $\chi
_{\bm{q},{\rm i},\omega}^{<}$, we define $\bar{b}_{\bm{q},{\rm i},\omega}=\chi
_{\bm{q},{\rm i},\omega}^{<}/\left[2i{\rm Im}\chi_{\bm{q},{\rm i},\omega}^{\rm ret.}\right]$.
While $\langle I_S\rangle_{\rm 1t}$ has been discussed in the context of cold atoms~\cite{Sekino2020},
$\langle I_S\rangle_{\rm spin}$ is a crucial term for spin transport extensively discussed in the field of spintronics~\cite{Konig2003,Adachi2011,Ohnuma2014,Ohnuma2017,Tatara2017,MM2018,Inoue2017,Kato2019,Silaev1,Silaev2,Ominato1,Ominato2,Kato2020,Yama2021,Yamamoto2021,Oue2021}.
In particular, if we consider the case with $\mathcal{T}_{\bm{k},\bm{p},\sigma}\rightarrow 0$ where the single-particle state is localized {(e.g., spin model)}, $\langle I_S\rangle_{\rm spin}$ driven by magnon excitations becomes dominant.
Even in the itinerant (delocalized) case, the spin current in the repulsive Fermi gas ($g>0$) would also be dominated by $\langle I_S\rangle_{\rm spin}$
because of the divergent spin susceptibility~\cite{Recati,Sandri}, which is known as Stoner ferromagnetism~\cite{Massignan} and has been observed in recent experiments~\cite{Valtolina}.
{In this regard, the effective tunneling Hamiltonian are common for both attractive and repulsive cases and the response functions uniquely determine the value of the current.}

\noindent {\it Conclusion.---}
In this study, we derived the {microscopic} tunneling Hamiltonian 
{at a strongly correlated interface.}
The one-body tunneling and reflection, and the pair- and spin-tunneling terms naturally arise from a microscopic model when taking the appropriate separation of the wave functions.
We demonstrated that the anomalous current induced by pair- and spin-tunneling arises even in the linear-response regime. This result is natural in the sense that the strong-coupling BEC limit should be dominated by the tunneling processes of molecular bosons.
We discussed how the pair-tunneling-induced mass conductance $\mathcal{G}$ and Seebeck coefficient $\mathcal{S}$ behave under the strong-coupling ansatz and demonstrated the anomalous enhancement of $\mathcal{G}/\mathcal{S}$ as a result of the Bose--Einstein statics of molecules. 
In addition, we derived the spin current formula for repulsive Fermi gases in terms of the spin susceptibility.
Our systematic framework will be useful for understanding transport phenomena in ultracold atoms and will be applicable to other systems such as condensed-matter and nuclear systems.

\par
{HT and MM are grateful to} Y.~Sekino, S.~Uchino, and Y.~Ominato for useful discussion.
HT also thanks Y. Guo for reading the manuscript.
The authors thank RIKEN iTHEMS NEW working group for fruitful discussions.
HT was supported by Grants-in-Aid for Scientific Research {provided by} JSPS {through No.~18H05406.}
DO is funded by the President’s PhD Scholarships at Imperial College London.
MM was supported by JSPS KAKENHI Grant Numbers JP20H01863 and JP21H04565, and the Priority Program of Chinese Academy of Sciences, under Grant No.~XDB28000000.

\begin{widetext}
\section{Decomposition of the wave function}
Let us consider scattering by one of the simple potentials,
\begin{align}
  V = 
  \begin{cases}{}
    V_0 & (|x| < \xv),
    \\
    0 & (|x| > \xv),
  \end{cases}
\end{align}
where we omit the spin index for simplicity.
Since our potential is a simple function,
we can write the wave function in a piecewise manner.
When we input a particle from the left-hand side,
the resultant wave function is
\begin{align}
  \psil
  &=
  \psil^\In \Big(
  \theta(-\xv-x) (e^{+ikx} + A_k e^{-ikx})
  +
  \theta(-x-\xv) B_k e^{+ikx}
  +\ldots
  \Big)
  :=
  \psil^\In f_{k\mathrm{L}}.
 \end{align}
As for the input from the right-hand side,
we have 
 \begin{align}
  \psir
  &=
  \psir^\In \Big(
  \theta(\xv-x)
  B_k e^{-ikx}
  +
  \theta(x-\xv)
  (e^{-ikx} + A_k e^{+ikx})
  +\ldots
  \Big)
  :=
  \psir^\In f_{k\mathrm{R}}.
\end{align}
Here,
we have defined the reflection and transmission coefficients,
$A_k$ and $B_k$.
We have also defined mode functions
$f_{k\mathrm{L}}$ and $f_{k\mathrm{R}}$.
They are mutually independent solutions to the Schr\"{o}dinger equation with the potential $V$ and obtained by fixing the input amplitudes,
$\psil^\In$ and $\psir^\In$,
at $x \rightarrow \pm\infty$.
Note that we have omitted the expression in the potential region ($|x| < \xv$).

After solving the scattering problem to find the reflection and transmission coefficients,
we can explicitly write the mode functions and can write the general solition to the Schr\"{o}dinger equation in terms of the mode functions.
In other words,
we can expand the wave function as the linear combination of the mode functions,
\begin{align}
  \psi
  &= \psil^\In f_{k\mathrm{L}} + \psir^\In f_{k\mathrm{R}}.
  \label{eq:split_psi}
\end{align}
We replace the coefficients with fermionic operators to quantise the field,
e.g.,
$
  \psil^\In f_{k\mathrm{L}}
  \mapsto
  c_{k,\mathrm{L}} f_{k\mathrm{L}}
  =: \hat{\psil}.
$
{Note that,
for any integrable potential which is represented in terms of simple functions,
we can formally decompose the wave function as done in Eq.~\eqref{eq:split_psi}.
Thus, the formal decomposition employed in the main text is justified even in the presence of soft walls.}

\section{Tunneling Hamiltonian in spinless fermions with a one-body barrier}
{Here, we consider a spinless Fermi gas as another example.
It is relevant for polarized electron gases as well as single-component Fermi gas near the $p$-wave Feshbach resonance.
The total Hamiltonian in the real space reads
\begin{align}
    \hat{H}=\int d^3\bm{r}
    \hat{\psi}^\dag(\bm{r})
    \hat{h}(\bm{r})
    \hat{\psi}(\bm{r})
    +\frac{1}{2}\int d^3\bm{r}\int d^3\bm{r}'
    \hat{\psi}^\dag(\bm{r})\hat{\psi}^\dag(\bm{r}')\hat{U}(\bm{r}-\bm{r}')
    \hat{\psi}(\bm{r}')\hat{\psi}(\bm{r}),
\end{align}
where
\begin{align}
    \hat{h}(\bm{r})=-\frac{\nabla^2}{2m}+\hat{V}(\bm{r})
\end{align}
is the kinetic energy of a fermion with mass $m$ and $\hat{V}(\bm{r})$ is the one-body potential barrier.
In the spinless case, the two-body interaction $\hat{U}(\bm{r}-\bm{r}')$ is nonlocal because the contact-type coupling is prohibited by the Pauli's exclusion principle.
Introducing the decomposed operators
\begin{align}
    \hat{\psi}(\bm{r})=\hat{\psi}_{\rm L}(\bm{r})+\hat{\psi}_{\rm R}(\bm{r}),
\end{align}
we rewrite $\hat{H}$ as
\begin{align}
    \hat{H}=\hat{H}_{{\rm L}}+\hat{H}_{{\rm R}}
    +\hat{H}_{\rm 1r}+\hat{H}_{\rm 1t}+\hat{H}_{\rm pair}
    +\hat{H}_{\rm ind.},
\end{align}
where the reservoir Hamiltonian
\begin{align}
    \hat{H}_{\rm i=L,R}=\int d^3\bm{r}
    \hat{\psi}_{\rm i}^\dag(\bm{r})
    \left[-\frac{\nabla^2}{2m}\right]
    \hat{\psi}_{\rm i}(\bm{r})
    +\frac{1}{2}\int d^3\bm{r}\int d^3\bm{r}'
    \hat{\psi}_{\rm i}^\dag(\bm{r})
    \hat{\psi}_{\rm i}^\dag(\bm{r}')\hat{U}(\bm{r}-\bm{r}')
    \hat{\psi}_{\rm i}(\bm{r}')\hat{\psi}_{\rm i}(\bm{r}),
\end{align}
the one-body reflection term
\begin{align}
    \hat{H}_{\rm 1r}=\int d^3\bm{r}\sum_{\rm i}\hat{\psi}_{\rm i}^\dag(\bm{r})\hat{V}(\bm{r})\hat{\psi}_{\rm i}(\bm{r}),
\end{align}
and the local one-body tunneling term
\begin{align}
    \hat{H}_{\rm 1t}=\int d^3\bm{r}
    \left[\hat{\psi}_{\rm L}^\dag(\bm{r})\hat{\tau}(\bm{r})\hat{\psi}_{\rm R}(\bm{r})+{\rm h.c.}\right],
\end{align}
are essentially same with the spin-$1/2$ case shown in the main text.
The one-body tunneling amplitude $\hat{\tau}(\bm{r})$ given by
\begin{align}
    \hat{\tau}(\bm{r})=\hat{h}(\bm{r})-\hat{U}(\bm{0})+\int d^3\bm{r}'\sum_{\rm i}\hat{U}(\bm{r}-\bm{r}')\hat{N}_{\rm i}(\bm{r}'),
\end{align}
reflects the nonlocal aspect of $\hat{U}(\bm{r}-\bm{r}')$.
Note that we have used $\hat{U}(\bm{r}-\bm{r}')=\hat{U}(\bm{r}'-\bm{r})$.
}

{The two-body tunneling term is given by
the pair-tunneling Hamiltonian
\begin{align}
    \hat{H}_{\rm pair}=\frac{1}{2}\int d^3\bm{r}d^3\bm{r}'
    \left[\hat{P}_{\rm L}^\dag(\bm{r},\bm{r}')\hat{U}(\bm{r}-\bm{r}')\hat{P}_{\rm R}(\bm{r},\bm{r}')+{\rm h.c.}\right]
\end{align}
where we defined the triplet-pair operator
$\hat{P}_{\rm i}(\bm{r},\bm{r}')=\psi_{\rm i}(\bm{r}')\psi_{\rm i}(\bm{r})$.
In this way, one can expect the triplet-pair transport in strongly interacting spinless fermionic systems.
Finally, we get two kinds of the induced interactions between reservoirs as $H_{\rm ind.}=H_{\rm dir.}+H_{\rm exc.}$.
The direct term $H_{\rm dir.}$ and the exchange term $H_{\rm exc.}$ are given by
\begin{align}
    \hat{H}_{\rm dir.}&=\int d^3\bm{r}d^3\bm{r}'
    \hat{C}_{\rm L}(\bm{r},\bm{r})\hat{U}(\bm{r}-\bm{r}')
    \hat{C}_{\rm R}(\bm{r}',\bm{r}') \\
    \hat{H}_{\rm exc.}&=-\frac{1}{2}\int d^3\bm{r}d^3\bm{r}'
    \left[\hat{C}_{\rm L}(\bm{r},\bm{r}')\hat{U}(\bm{r}-\bm{r}')\hat{C}_{\rm R}(\bm{r}',\bm{r})+{\rm h.c.}\right],
\end{align}
respectively, where we defined the generalized density operator $\hat{C}_{\rm i}(\bm{r},\bm{r}')=\hat{\psi}_{\rm i}^\dag(\bm{r})\hat{\psi}_{\rm i}(\bm{r}')$.
The appearance of $H_{\rm exc.}$ is in contrast to the case with the contact-type interaction between different spins.
These inter-reservoir interaction can be important when considering inhomogenous reservoirs.
}

{Taking the same procedure with the spin-$1/2$ case, we can obtain the effective tunneling Hamiltonian in the momentum space.
In particular,
the triplet-pair tunneling term is approximately given by
\begin{align}
    H_{\rm pair}\simeq\frac{1}{2}\sum_{\bm{k},\bm{k}',\bm{q}}\left[\mathcal{T}_{2,{\rm pair},}(\bm{k},\bm{k}',\bm{q})c_{\bm{k}+\bm{q}/2,{\rm L}}^\dag c_{-\bm{k}+\bm{q}/2,{\rm L}}^\dag
    c_{-\bm{k}'-\bm{q}/2,{\rm R}}
    c_{\bm{k}'-\bm{q}/2,{\rm R}}
    +{\rm h.c.}\right],
\end{align}
where
$\mathcal{T}_{2,{\rm pair},}(\bm{k},\bm{k}',\bm{q})
=U_{-\bm{k}-\bm{k}'}[B_{\bm{k}+\bm{q}/2}^*B_{-\bm{k}+\bm{q}/2}^*+B_{-\bm{k}'-\bm{q}/2}B_{\bm{k}'-\bm{q}/2}]$.
$B_{\bm{k}}$ is the one-body tunneling amplitude with respect to $\hat{V}(\bm{r})$.
}

{
Furthermore, we briefly mention the three-body tunneling process induced by the three-body interaction
\begin{align}
    \hat{W}=\frac{1}{6}\int d^3\bm{r}
    \int d^3\bm{r}' \int d^3\bm{r}''
    \hat{\psi}^\dag(\bm{r})
    \hat{\psi}^\dag(\bm{r}')
    \hat{\psi}^\dag(\bm{r}'')
    \hat{U}_3(\bm{r},\bm{r}',\bm{r}'')
    \hat{\psi}(\bm{r}'')
    \hat{\psi}(\bm{r}')
    \hat{\psi}(\bm{r}),
\end{align}
with the coupling strength $\hat{U}_3(\bm{r},\bm{r}',\bm{r}'')$.
Again, we can use decomposed operator $\hat{\psi}(\bm{r})=\hat{\psi}_{\rm L}(\bm{r})+\hat{\psi}_{\rm R}(\bm{r})$ and find the three-body tunneling Hamiltonian in the momentum space as
\begin{align}
    H_{\rm triple}\simeq\frac{1}{6}
    \sum_{\bm{k}_{1},\bm{k}_{2},\bm{k}_{3},\bm{q}_1,\bm{q}_2}
    \left[\mathcal{T}_3(\bm{k}_1,\bm{k}_2,\bm{k}_3,\bm{q}_1,\bm{q}_2)
    c_{\bm{k}_1,{\rm L}}^\dag c_{\bm{k}_2, {\rm L}}^\dag c_{\bm{k}_3, {\rm L}}^\dag
    c_{-\bm{k}_3+\bm{q}_2, {\rm R}} c_{-\bm{k}_2+\bm{q}_1-\bm{q}_2, {\rm R}} c_{-\bm{k}_1-\bm{q}_1, {\rm R}}
    +{\rm h.c.}\right],
\end{align}  
where we have defined $\mathcal{T}_3(\bm{k}_1,\bm{k}_2,\bm{k}_3,\bm{q}_1,\bm{q}_2)=U_3(\bm{q}_1,\bm{q}_2)\left[B_{\bm{k}_1}^*B_{\bm{k}_2}^*B_{\bm{k}_3}^*
+B_{-\bm{k}_3+\bm{q}_2}B_{-\bm{k}_2+\bm{q}_1-\bm{q}_2}B_{-\bm{k}_1-\bm{q}_1}
\right]$ and used the Fourier transformation $U_3(\bm{r},\bm{r}',\bm{r}'')=\sum_{\bm{q}_1,\bm{q}_2}U_3(\bm{q}_1,\bm{q}_2)e^{-i\bm{q}_1\cdot(\bm{r}-\bm{r}')-i\bm{q}_2\cdot(\bm{r}'-\bm{r}'')}$.
In this way, if the bulk system has the non-negligible three-body interactions, one may expect three-particle tunneling process induced by the three-body force.
Our framework can be extended to cases with more than three-body interactions.
}

\section{Pair-tunneling-induced mass conductance and Seebeck coefficient within the strong-coupling ansatz}
{The molecular number density $N_{\rm mol.}$ can be estimated from
\begin{align}
    N_{\rm mol.}=-2\sum_{\bm{q}}\int_{-\infty}^{\infty}\frac{d\omega}{\pi}\frac{{\rm Im}D_{\bm{q},\omega}^{\rm ret.}}{e^{\frac{\omega-2\mu}{T}}-1},
\end{align}
where
\begin{align}
    D_{\bm{q},\omega}^{\rm ret.}=\frac{1}{\omega+i\delta-\frac{q^2}{4m}+2\mu-E_{\rm b}-\Sigma_{\bm{q},\omega}}
\end{align}
is the retarded Green's function of a molecular boson with the self-energy $\Sigma_{\bm{q},\omega}$.
$E_{\rm b}=1/(ma^2)$
is the two-body binding energy.
Here we suppressed the index $i$ for the reservoirs.
For simplicity, we have employed a phenomenological self-energy $\Sigma_{\bm{q},\omega}\simeq\Sigma_{0}-i\gamma\theta(\omega)$, 
where the damping $\gamma=0.1E_{\rm F}$ is associated with multiple boson--boson scatterings.
The constant shift $\Sigma_0$ can be absorbed into the shift of $E_{\rm b}$ as $\tilde{E}_{\rm b}=E_{\rm b}+\Sigma_0$. 
While the self-energy in weakly-repulsive Bose gases have been extensively discussed in connection with density fluctuations~\cite{Watabe2021},
these details are out of scope in this work.
In the practical calculations, $\mu$ is obtained by the strong-coupling ansatz
\begin{align}
N_{{\rm tot.}}\simeq N_{\rm mol.}\simeq2\sum_{\bm{q}}\frac{1}{e^{\frac{q^2/4m-2\mu+\tilde{E}_{\rm b}}{T}}-1}.
\end{align}
Note that we approximately obtain $\Gamma_{\bm{q},\omega}^{\rm ret.}\simeq ZD_{\bm{q},\omega}^{\rm ret.}$ in the strong-coupling limit, where the wave function renormalization $Z$ involves the ultraviolet cutoff~\cite{Tajima2022}.}
The fermion number density is strongly suppressed by the emergence of tightly bound molecules in the strong-coupling regime, $1/(k_{\rm F}a)\gesim 1$.

We numerically confirmed that a value of $\gamma_i$ does not qualitatively change our result.
While the value of $a^{-1}$ modifies the coefficients of $\mathcal{G}$ and $\mathcal{S}$ associated with $g^2$ and $Z$, the ratio $\mathcal{G}/\mathcal{S}$ does not explicitly depend on them in the limit of $a^{-1}\rightarrow \infty$.
{While we keep $\gamma$ to be finite for the numerical calculation,
such a treatment is not necessary for the case with momentum-unconserved tunneling coupling.
}

\end{widetext}

\end{document}